\documentclass[a4paper, 10 pt, conference]{ieeeconf}
\IEEEoverridecommandlockouts
\usepackage{amsfonts,amsmath,amssymb} % assumes amsmath package installed

\usepackage{psfrag,color}
\usepackage{enumerate,cite,latexsym,graphicx}
\newtheorem{theorem}{Theorem}
\newtheorem{lemma}{Lemma}

\newtheorem{definition}{Definition}

\def\tr{\mathop{\rm Tr}\nolimits} % Trace of matrix

\title{\LARGE \bf Robust Stability of  Quantum Systems with Nonlinear Dynamic Uncertainties
}

\author{Ian R.~Petersen %
\thanks{This work was supported by the
Australian Research Council (ARC) and the Air Force Office of Scientific
Research (AFOSR). This material is based on research sponsored by the
Air Force Research Laboratory, under agreement numbers
FA2386-09-1-4089 and FA2386-12-1-4075.  The U.S. Government is authorized to reproduce and
distribute reprints for Governmental purposes notwithstanding any
copyright notation thereon.
The views and conclusions contained herein are those of the authors
and should not be interpreted as necessarily representing the official
policies or endorsements, either expressed or implied, of the Air
Force Research Laboratory or the U.S. Government. }%
\thanks{Ian R. Petersen is with the School of  Engineering and Information Technology, 
        University of New South Wales at the Australian Defence Force Academy, Canberra ACT 2600, Australia.
         {\tt\small i.r.petersen@gmail.com} } 
}%

\begin{document}

\maketitle
\thispagestyle{empty}
\pagestyle{empty}

\begin{abstract}
This paper considers the problem of robust stability for a class of
uncertain nonlinear quantum systems subject to unknown perturbations in the
system Hamiltonian. The nominal system is a linear quantum system defined by a linear vector of coupling operators and a quadratic Hamiltonian. 
This paper extends previous results on the robust stability of nonlinear quantum systems to allow for quantum systems with dynamic uncertainties. These dynamic uncertainties are required to satisfy a certain quantum stochastic integral quadratic constraint. The robust stability condition is given in terms of a strict bounded real condition. This result is applied  to the robust stability analysis of  an optical parametric amplifier.
\end{abstract}

%%%%%%%%%%%%%%%%%%%%%%%%%%%%%%%%%%%%%%%%%%%%%%%%%%%%%%%%%%%%%%%%%%%%%%%%%%%%%%%%
\section{Introduction} \label{sec:intro}
In recent years, a number of papers have considered the feedback
control of systems whose dynamics are governed by the laws of quantum 
mechanics instead of classical mechanics; see e.g., \cite{YK03A,YK03B,YAM06,JNP1,NJP1,GGY08,MaP3,MaP4,YNJP1,GJ09, GJN10,WM10,PET10Ba}. In particular, the papers \cite{GJ09,JG10} consider a framework for
quantum systems defined in terms of a triple $(S,L,H)$ where $S$ is a
Scattering Matrix,  $L$ is a vector of  coupling operators, and $H$ is a
Hamiltonian operator.

The papers \cite{PUJ1a, PUJ2} consider the problem of absolute stability of a quantum system defined in terms of a triple $(S,L,H)$ where the quantum system Hamiltonian is
decomposed as $H = H_1 + H_2$, $H_1$ is a known nominal Hamiltonian
and $H_2$ a perturbation Hamiltonian, which is contained a
specified set of Hamiltonians $\mathcal {W} $. In particular, the papers \cite{PUJ1a,PUJ2}  consider the case in which the 
nominal Hamiltonian $H_1$ is a quadratic function of  annihilation and
creation operators and the elements of the coupling operator vector are linear 
functions of the annihilation and creation operators. This case corresponds to
a nominal linear quantum system; for example, see
\cite{JNP1,NJP1,MaP3,MaP4,PET10Ba}. The results  in \cite{PUJ1a,PUJ2} were extended in \cite{PUJ3a} to allow for uncertainty in the coupling operator $L$.  Also, the results of \cite{PUJ1a} have been used in the robust stability analysis of a quantum system
consisting of a Josephson junction in a resonant cavity; see \cite{PET12Aa}.

In the paper  \cite{PUJ1a}, it is assumed that $ H_2$ is contained in a set of non-quadratic perturbation Hamiltonians bounded according to a sector bound on the nonlinearity. In this case, \cite{PUJ1a}  obtained a robust stability result in terms of a frequency domain condition. This result can be regarded as a quantum version of the classical small gain theorem for the case of static sector bounded nonlinearities; e.g., see \cite{KHA02}. Also, the paper \cite{PUJ2}  limited attention to quadratic perturbation Hamiltonians. In this case, a frequency domain robust stability condition is also obtained.

It is well known that the classical small gain robust stability condition also applies in the case of nonlinear dynamic uncertainties. Such uncertainties can be described in terms of integral quadratic constraints (IQCs); e.g., see \cite{PUSB}. The main result of this paper extends the quantum small gain result of \cite{PUJ1a} to allow for nonlinear dynamic uncertainties which are described by a certain quantum stochastic integral quadratic constraint (QSIQC). This uncertainty description can be regarded as a continuous time quantum version of the stochastic uncertainty constraint considered in \cite{PJ1}. In our main result, the presence of dynamic uncertainties is represented by a perturbation Hamiltonian which depends on system variables which are in addition to those which occur in the nominal system Hamiltonian. 

An example in the paper  \cite{PUJ2}  considers the robust stability analysis of a quantum system consisting of a linearized optical parametric amplifier (OPA). Optical parametric amplifiers are widely used in the field of experimental quantum optics used; see e.g. \cite{GZ00, BR04, WM08}. In particular, they can be used to generate squeezed light, which has a smaller noise variance in one quadrature than the standard quantum limit would allow. This is at the expense of a larger noise variance in the other quadrature; see e.g. \cite{GZ00,BR04,WM08,SHHPJ2a,SaP2a,SaPH1a}. Such an OPA can be made using a nonlinear optical medium in an optical cavity; for example, see \cite {BR04,SHHPJ2a,SaP2a,SaPH1a}. This allows for the interaction between a fundamental optical field and a second harmonic optical field. The paper \cite {PUJ2} analyzed  a linearized model of the OPA which considered only the fundamental field and the fundamental mode of the cavity. To illustrate the results of this paper, we will analyze a linearized model of an OPA which considers both the fundamental and second harmonic fields and cavity modes. In this case, the second harmonic cavity mode will be considered as a dynamic uncertainty satisfying a QSIQC.

\section{Quantum Systems with Nonlinear Dynamic Uncertainties} \label{sec:systems}
In this section, we describe the general class of quantum systems under consideration. 
As in the papers \cite{GJ09,JG10,PUJ1a,PUJ2,JPU1a},  we consider uncertain nonlinear open quantum systems defined by  parameters $(S,L,H)$ where $S$ is the scattering matrix, which is typically chosen as the identity matrix, L is the coupling operator and $H$ is the system  Hamiltonian operator which is assumed to be of the form
\begin{equation}
\label{H1}
H = \frac{1}{2}\left[\begin{array}{cc}a^\dagger &
      a^T\end{array}\right]M
\left[\begin{array}{c}a \\ a^\#\end{array}\right]+f(b,b^\#,z,z^*).
\end{equation}
Here $a$ is a vector of annihilation
operators on the underlying Hilbert space and $a^\#$ is the
corresponding vector of creation operators. Also the vectors of operators, $b$ and $b^\#$ are defined similarly. Furthermore,  $M \in \mathbb{C}^{2n\times 2n}$ is a Hermitian matrix of the
form
\begin{equation}
\label{Mform}
M= \left[\begin{array}{cc}M_1 & M_2\\
M_2^\# &     M_1^\#\end{array}\right]
\end{equation}
and $M_1 = M_1^\dagger$, $M_2 = M_2^T$.
In the case of vectors of
operators, the notation  $^\dagger$ refers to the transpose of the vector of adjoint
operators and  in the case of matrices, this notation refers to the complex conjugate transpose of a matrix. In the case vectors of
operators, the notation $^\#$ refers to the vector of adjoint
operators and in the case of complex matrices, this notation refers to
the complex conjugate matrix. Also, the notation $^*$ denotes the adjoint of an
operator. The matrix $M$ is assumed to be known and defines the nominal quadratic part of the system Hamiltonian. 
  Furthermore, we assume the uncertain non-quadratic  part of the system Hamiltonian  $f(b,b^\#,z,z^*)$ is defined by a formal power series of  the form
\begin{eqnarray}
\label{H2nonquad}
f(b,b^\#,z,z^*)
&=& \sum_{k=0}^\infty\sum_{\ell=0}^\infty S_{k\ell}(b,b^\#)z^k(z^*)^\ell\nonumber \\
&=& \sum_{k=0}^\infty\sum_{\ell=0}^\infty S_{k\ell}(b,b^\#)H_{k\ell},
\end{eqnarray}
which is assumed to converge in some suitable sense.
Here $S_{k\ell}(b,b^\#)=S_{\ell k}(b,b^\#)^*$, $H_{k\ell} = z^k(z^*)^\ell$,  and $z$ is a known scalar operator defined by
\begin{eqnarray}
\label{z}
z &=&  E_1a+E_2 a^\# \nonumber \\
&=& \left[\begin{array}{cc} E_1 & E_2 \end{array}\right]
\left[\begin{array}{c}a \\ a^\#\end{array}\right] = \tilde E 
\left[\begin{array}{c}a \\ a^\#\end{array}\right].
\end{eqnarray}

The term $f(b,b^\#,z,z^*)$ is referred to as the perturbation Hamiltonian. It  is assumed to be unknown but is contained within a known set which will be defined below.
It follows from this definition  that
 $f(b,b^\#,z,z^*)$ is a self-adjoint operator. The fact that $f(b,b^\#,z,z^*)$ depends on the quantities $b$ and $b^\#$ which do not appear in the nominal Hamiltonian corresponds to our assumption that we allow nonlinear dynamic uncertainties in the quantum system. 

We assume the coupling operator $L$ is known and is of the form 
\begin{equation}
\label{L}
L = \left[\begin{array}{c}
L_a\\L_b
\end{array}\right]
\end{equation}
where 
\begin{equation}
\label{La}
L_a = \left[\begin{array}{cc}N_{a1} & N_{a2}\end{array}\right]\left[\begin{array}{c}a \\ a^\#\end{array}\right]
\end{equation}
and 
\begin{equation}
\label{Lb}
L_a =\left[\begin{array}{cc} N_{b1} &  N_{b2}\end{array}\right]
\left[\begin{array}{c}b \\b^\#\end{array}\right].
\end{equation}
Here,  $N_{a1} \in \mathbb{C}^{m\times n}$,  $N_{a2} \in
\mathbb{C}^{m\times n}$, $ N_{b1} \in \mathbb{C}^{\tilde m\times \tilde n}$ and  $ N_{b2} \in
\mathbb{C}^{\tilde m\times \tilde n}$. Also, we write
\begin{eqnarray*}
\left[\begin{array}{c}L_a \\ L_a^\#\end{array}\right] &=& N_a
\left[\begin{array}{c}a \\ a^\#\end{array}\right] \\
&=&
\left[\begin{array}{cc}N_{a1} & N_{a2}\\
N_{a2}^\# &     N_{a1}^\#\end{array}\right] 
\left[\begin{array}{c}a \\ a^\#\end{array}\right]
\end{eqnarray*} 
and
\begin{eqnarray*}
\left[\begin{array}{c}L_b \\ L_b^\#\end{array}\right]
&=& N_b \left[\begin{array}{c}b \\ b^\#\end{array}\right]\\
&=&\left[\begin{array}{cc} N_{b1} &  N_{b2}\\
 N_{b2}^\# &      N_{b1}^\#\end{array}\right]\left[\begin{array}{c}b \\ b^\#\end{array}\right].
\end{eqnarray*} 

The annihilation and creation operators $a$ and $a^\#$ are assumed to satisfy the
canonical commutation relations:
\begin{eqnarray}
\label{CCR2}
\left[\left[\begin{array}{l}
      a\\a^\#\end{array}\right],\left[\begin{array}{l}
      a\\a^\#\end{array}\right]^\dagger\right]
&\stackrel{\Delta}{=}&\left[\begin{array}{l} a\\a^\#\end{array}\right]
\left[\begin{array}{l} a\\a^\#\end{array}\right]^\dagger
\nonumber \\
&&- \left(\left[\begin{array}{l} a\\a^\#\end{array}\right]^\#
\left[\begin{array}{l} a\\a^\#\end{array}\right]^T\right)^T\nonumber \\
&=& J
\end{eqnarray}
where $J = \left[\begin{array}{cc}I & 0\\
0 & -I\end{array}\right]$; e.g., see \cite{GGY08,GJN10,PET10Ba}. Similarly, we assume
\begin{eqnarray}
\label{CCR2b}
\left[\left[\begin{array}{l}
      b\\b^\#\end{array}\right],\left[\begin{array}{l}
      b\\b^\#\end{array}\right]^\dagger\right] = J.
\end{eqnarray}
Also, we assume that all of the elements of the vectors $a$ and $a^\#$ commute with all of the elements of the vectors $b$ and $b^\#$.

To define the set of allowable perturbation Hamiltonians $f(\cdot)$, we first define the following formal partial derivatives:
\begin{equation}
\label{fdash}
\frac{\partial f(b,b^\#,z,z^*)}{\partial z} \stackrel{\Delta}{=} \sum_{k=1}^\infty\sum_{\ell=0}^\infty k S_{k \ell}(b,b^\#)z^{k-1}(z^*)^\ell;
\end{equation}
\begin{equation}
\label{fddash}
\frac{\partial^2 f(b,b^\#,z,z^*)}{\partial z^2} 
\stackrel{\Delta}{=} \sum_{k=1}^\infty\sum_{\ell=0}^\infty k(k-1)S_{k\ell}(b,b^\#) z^{k-2}(z^*)^{\ell}.
\end{equation}
Then, we consider the following quantum stochastic differential equations describing the uncertainty dynamics (e.g., see equations (1) and (2) in \cite{EMPUJ3a} and equations (7) and (9) in \cite{ShP5}):
\begin{eqnarray}
\label{uncert_dynamics}
d \left[\begin{array}{l}
      b\\b^\#\end{array}\right] &=&
-\imath\left[ 
\left[\begin{array}{l}b\\b^\#\end{array}\right]
,f(b,b^\#,z,z^*)\right]dt \nonumber \\
&&+\frac{1}{2}\left(L_b^\dagger\left[ 
\left[\begin{array}{l} b\\b^\#\end{array}\right]
,L_b^T\right]^T\right)^Tdt \nonumber \\
&& +\frac{1}{2}\left[L_b^\#,
\left[\begin{array}{l} b\\b^\#\end{array}\right]^T
\right]^TL_bdt \nonumber \\
&& + \left[
\left[\begin{array}{l}b\\b^\#\end{array}\right]
,L^T\right]d\mathcal{B}^\#\nonumber \\
&&- \left[
\left[\begin{array}{l}b\\b^\#\end{array}\right]
,L^\dagger\right]d\mathcal{B}\nonumber \\
&=& -\imath \sum_{k=0}^\infty\sum_{\ell=0}^\infty \left[
\left[\begin{array}{l}b\\b^\#\end{array}\right]
,S_{k\ell}(b,b^\#)\right]z^k(z^*)^\ell dt \nonumber \\
&&-\frac{1}{2}JN_b^\dagger J N_b 
\left[\begin{array}{l}b\\b^\#\end{array}\right] dt\nonumber \\
&&-JN_b^\dagger J 
\left[\begin{array}{l}d\mathcal{B}\\d\mathcal{B}^\#\end{array}\right],
\end{eqnarray}
where $\mathcal{B}(t)$ is a vector of bosonic annihilation operators corresponding to the quantum fields acting on the uncertainty system and $\mathcal{B}(t)^\#$ is the corresponding vector of creation operators; e.g., see \cite{PAR92}. The vector $\mathcal{B}(t)$ corresponds to a vector of standard quantum Weiner processes.  The set of allowable perturbation Hamiltonians will be defined in terms of quantum stochastic integral quadratic constraints (QSIQCs) for the system (\ref{uncert_dynamics}). These conditions are defined in a similar way to the definition of dissipativity in \cite{JG10}; i.e., the given inequalities are required to hold for all interconnections between the system 
(\ref{uncert_dynamics}) and an exosystem $\tilde W$ contained in a suitable class of exosystems $\mathcal{\tilde W}$.

 For given constants $\gamma > 0$ and  $\delta_1\geq 0$,  we consider the 
QSIQC 
\begin{equation}
\label{sector4a}
\limsup_{T\rightarrow \infty} \frac{1}{T}\int_0^T\left(\left<w_1(t) w_1(t)^*\right>-  \frac{1}{\gamma^2}\left<z(t) z(t)^*\right>\right)dt \leq \delta_1,
\end{equation}
where 
\[
w_1(t) = \frac{\partial f(b(t),b(t)^\#,z(t),z(t)^*)^*}{\partial z}
\]
and $b(t)$, $b(t)^\#$, $z(t)$, $z(t)^*$ denote the Heisenberg evolutions of the operators $b$, $b^\#$, $z$, $z^*$
respectively for the system formed by the interconnection between the quantum system (\ref{uncert_dynamics}) and an exosystem $\tilde W$; e.g., see \cite{JG10}.
Similarly, for a  given constant
$\delta_2\geq 0$, we consider the 
QSIQC 
\begin{equation}
\label{sector4b}
\limsup_{T\rightarrow \infty} \frac{1}{T}\int_0^T\left<w_2(t) w_2(t)^*\right>dt \leq  \delta_2,
\end{equation}
where
\[
w_2(t) = \frac{\partial^2f(z,z^*)^*}{\partial z^2}.
\]
Here the notation $\left<\cdot\right>$ denotes quantum expectation; e.g., see \cite{PAR92}.
Then we define the set of perturbation Hamiltonians $\mathcal{W}$  as follows:
\begin{equation}
\label{W5}
\mathcal{W} = \left\{\begin{array}{l}f(\cdot) \mbox{ of the form
      (\ref{H2nonquad}) such that 
} \\
\mbox{ conditions (\ref{sector4a}) and (\ref{sector4b}) are satisfied}\end{array}\right\}.
\end{equation}

We will consider the following notion of robust mean square stability which is somewhat different from the definition considered in \cite{PUJ1a} due to the presence of dynamic uncertainties. 
\begin{definition}
\label{D1}
An uncertain open quantum system defined by  $(S,L,H)$ where $H$ is of the form (\ref{H1}), $f(\cdot) \in \mathcal{W}$, and $L$ is of the form (\ref{L}) is said to be {\em robustly mean square stable} if there exists a constant $c > 0$,  such that for any $f(\cdot) \in \mathcal{W}$,  
\begin{eqnarray}
\label{ms_stable0}
\limsup_{T\rightarrow \infty} \frac{1}{T}\int_0^T\left< \left[\begin{array}{c}a(t) \\ a^\#(t)\end{array}\right]^\dagger \left[\begin{array}{c}a(t) \\ a^\#(t)\end{array}\right] \right>dt&\leq& c.\nonumber \\
\end{eqnarray}
Here $\left[\begin{array}{c}a(t) \\ a^\#(t)\end{array}\right]$ denotes the Heisenberg evolution of the vector of operators $\left[\begin{array}{c}a \\ a^\#\end{array}\right]$.
\end{definition}
\section{Main Results}
\label{sec:main}
We will show that  the following small gain condition
is sufficient for the robust mean square stability
of the nonlinear quantum system under consideration when $f(\cdot) \in \mathcal{W}$: 
\begin{enumerate}
\item
The matrix 
\begin{equation}
\label{Hurwitz1}
F = -\imath JM-\frac{1}{2}JN_a^\dagger J N_a\mbox{ is Hurwitz;}
\end{equation}
\item
\begin{equation}
\label{Hinfbound1}
\left\|\tilde E^\# \Sigma\left(sI -F\right)^{-1}J\Sigma \tilde E^T \right\|_\infty < \frac{\gamma}{2}
\end{equation}
where  $\Sigma = \left[\begin{array}{cc} 0 & I\\
I &0 \end{array}\right].
$
\end{enumerate}
This leads to the following theorem.

\begin{theorem}
\label{T1}
Consider an uncertain open nonlinear quantum system defined by $(S,L,H)$  such that
$S=I$, $H$ is of the form (\ref{H1}), $L$ is of the
form (\ref{L}) and $f(\cdot) \in \mathcal{W}$. Furthermore, assume that
the strict bounded real condition  (\ref{Hurwitz1}), (\ref{Hinfbound1})
is satisfied. Then the
uncertain quantum system is robustly mean square stable.
\end{theorem}

To prove this theorem, we will consider quadratic ``Lyapunov'' operators  $V$ of the form 
\begin{equation}
\label{quadV}
V = \left[\begin{array}{cc}a^\dagger &
      a^T\end{array}\right]P
\left[\begin{array}{c}a \\ a^\#\end{array}\right]
\end{equation}
where $P \in \mathbb{C}^{2n\times 2n}$ is a positive-definite Hermitian matrix of the
form
\begin{equation}
\label{Pform}
P= \left[\begin{array}{cc}P_1 & P_2\\
P_2^\# &     P_1^\#\end{array}\right].
\end{equation}
 Hence, we consider a set of  non-negative self-adjoint operators
$\mathcal{P}$ defined as
\begin{equation}
\label{P1}
\mathcal{P} = \left\{\begin{array}{l}V \mbox{ of the form
      (\ref{quadV}) such that $P > 0$ is a 
} \\
\mbox{  Hermitian matrix of the form (\ref{Pform})}\end{array}\right\}.
\end{equation}

\begin{lemma}
\label{L4}
Given any $V \in \mathcal{P}$, then
\begin{equation}
\label{mui}
\mu = \left[z,[z,V]\right] = \left[z^*,[z^*,V]\right]^* = 
-\tilde E \Sigma JPJ\tilde E^T,
\end{equation}
which is a constant.  
\end{lemma}
{\em Proof:}
The proof of this result follows via a straightforward but tedious
calculation using (\ref{CCR2}). \hfill $\Box$

\begin{lemma}
\label{LB}
Given any $V \in \mathcal{P}$, then
\begin{eqnarray}
\label{comm_condition}
[V,f(b,b^\#,z,z^*)] &=&[V,z ]w_{1}^* -w_{1}[z^*,V]\nonumber \\
&&+ \frac{1}{2}\mu  w_{2}^*-\frac{1}{2}w_{2}\mu^*
\end{eqnarray} 
where
\begin{eqnarray}
\label{zw1w2}
w_1&=&  = \frac{\partial f((b,b^\#,z,z^*)}{\partial z }^*,\nonumber \\
w_2&=&  = \frac{\partial^2 f((b,b^\#,z,z^*)}{\partial z ^2}^*,\nonumber \\
\end{eqnarray}
and the constant $\mu $ is defined as in (\ref{mui}). 
\end{lemma}

\noindent
{\em Proof:}
First, we note that given any $V \in \mathcal{P}$,  and $k \geq 1$,
\begin{eqnarray}
\label{Vzetak}
 Vz   &=& [V,z ]+ z  V;\nonumber \\
\vdots && \nonumber \\
Vz ^k &=& \sum_{n=1}^k z ^{n-1}[V,z ] z ^{k-n}+z ^k V.
\end{eqnarray}
Also using Lemma \ref{L4}, it follows that for any $n \geq 1$,
\begin{eqnarray}
\label{Vzetak1}
z  [V,z ] &=& [V,z ]z  + \mu ; \nonumber \\
\vdots && \nonumber \\
 z ^{n-1} [V,z ]&=&[V,z ]z ^{n-1}
+ (n-1)z ^{n-2}\mu .
\end{eqnarray}
Therefore using (\ref{Vzetak}) and (\ref{Vzetak1}), it follows that
\begin{eqnarray*}
Vz ^k &=&\sum_{n=1}^k [V,z ] z ^{n-1}z ^{k-n}+ (n-1)z ^{n-2}z ^{k-n}\mu   \nonumber \\
&&+z ^k V\nonumber \\
&=& \sum_{n=1}^k [V,z ] z ^{k-1}+ (n-1)z ^{k-2}\mu   +z ^k V\nonumber \\
&=&k[V,z ] z ^{k-1}+\frac{k(k-1)}{2}z ^{k-2}\mu +z ^k V 
\end{eqnarray*}
which holds for any  $k \geq 0$. 
Similarly for any  $\ell \geq 0$,
\begin{eqnarray*}
(z^*)^\ell V &=& \ell(z^*)^{\ell-1}[z^*,V]+\frac{\ell(\ell-1)}{2}\mu^*(z^*)^{\ell-2}
\nonumber \\
&&+V(z^*)^\ell.
\end{eqnarray*}

Now given any $k \geq 0$, $\ell \geq 0$,  we have using the notation in (\ref{H2nonquad}):
\begin{eqnarray}
\label{VHkl}
\lefteqn{[V,H_{k\ell}]}\nonumber \\
 &=& k[V,z ] z ^{k-1}(z^*)^\ell+\frac{k(k-1)}{2}\mu z ^{k-2}(z^*)^\ell\nonumber \\&&
+z ^k V(z^*)^\ell\nonumber \\
&& -\ell z ^k(z^*)^{\ell-1}[z^*,V]-\frac{\ell(\ell-1)}{2}\mu^*z ^k(z^*)^{\ell-2}\nonumber \\&&
-z ^k V(z^*)^\ell\nonumber \\
&=& k[V,z ] z ^{k-1}(z^*)^\ell-\ell z ^k(z^*)^{\ell-1}[z^*,V]\nonumber \\&&
+\frac{k(k-1)}{2}\mu z ^{k-2}(z^*)^\ell-\frac{\ell(\ell-1)}{2}\mu^*z ^k(z^*)^{\ell-2}.\nonumber \\
\end{eqnarray}
Therefore,
\begin{eqnarray}
\label{VH2}
[V,f(b,b^\#,z,z^*)] &=& \sum_{k=0}^\infty \sum_{\ell=0}^\infty S_{k\ell}(b,b^\#) [V,H_{k\ell}] \nonumber \\
&=& [V,z ]\frac{\partial f(b,b^\#,z,z^*)}{\partial z } \nonumber \\
&&-\frac{\partial f(b,b^\#,z,z^*)}{\partial z}^*[z^*,V]\nonumber \\
&&+ \frac{1}{2}\mu  \frac{\partial^2 f(b,b^\#,z,z^*)}{\partial z^2}\nonumber \\
&&-\frac{1}{2}\frac{\partial^2 f(b,b^\#,z,z^*)}{\partial z^2}^*\mu^*.
\end{eqnarray}
Now 
 it follows from (\ref{zw1w2}) that condition (\ref{comm_condition}) is
 satisfied. 
\hfill $\Box$

\begin{lemma}
\label{L2}
Given $V \in \mathcal{P}$ and $L_a$ defined as in (\ref{L}), then
\begin{eqnarray*}
\lefteqn{[V,\frac{1}{2}\left[\begin{array}{cc}a^\dagger &
      a^T\end{array}\right]M
\left[\begin{array}{c}a \\ a^\#\end{array}\right]] =}\nonumber \\
&& \left[\left[\begin{array}{cc}a^\dagger &
      a^T\end{array}\right]P
\left[\begin{array}{c}a \\ a^\#\end{array}\right],\frac{1}{2}\left[\begin{array}{cc}a^\dagger &
      a^T\end{array}\right]M
\left[\begin{array}{c}a \\ a^\#\end{array}\right]\right] \nonumber \\
&=& \left[\begin{array}{c}a \\ a^\#\end{array}\right]^\dagger 
\left[
PJM - MJP 
\right] \left[\begin{array}{c}a \\ a^\#\end{array}\right].
\end{eqnarray*}
Also,
\begin{eqnarray*}
\lefteqn{\frac{1}{2}L_a^\dagger[V,L_a]+\frac{1}{2}[L_a^\dagger,V]L_a =} \nonumber \\
&=& \tr\left(PJN_a^\dagger\left[\begin{array}{cc}I & 0 \\ 0 & 0 \end{array}\right]N_aJ\right)
\nonumber \\&&
-\frac{1}{2}\left[\begin{array}{c}a \\ a^\#\end{array}\right]^\dagger
\left(N_a^\dagger J N_a JP+PJN_a^\dagger J N_a\right)
\left[\begin{array}{c}a \\ a^\#\end{array}\right].
\end{eqnarray*}
Furthermore, 
\[
\left[\left[\begin{array}{c}a \\ a^\#\end{array}\right],\left[\begin{array}{cc}a^\dagger &
      a^T\end{array}\right]P
\left[\begin{array}{c}a \\ a^\#\end{array}\right]\right] = 2JP\left[\begin{array}{c}a \\ a^\#\end{array}\right].
\]
\end{lemma}
{\em Proof:}
The proof of these identities follows via  straightforward but tedious
calculations using (\ref{CCR2}). \hfill $\Box$

\noindent
{\em Proof of Theorem \ref{T1}.}
It follows from (\ref{z}) that we can write
\begin{eqnarray*}
z^* &=& E_1^\#a^\#+E_2^\# a=\left[\begin{array}{cc} E_2^\# & E_1^\# \end{array}\right]
\left[\begin{array}{c}a \\ a^\#\end{array}\right]\nonumber \\
&=&  \tilde E^\# \Sigma \left[\begin{array}{c}a \\ a^\#\end{array}\right].
\end{eqnarray*}
Also,  it follows from Lemma \ref{L2} that
\[
[z^*,V] = 2 \tilde E^\# \Sigma
JP\left[\begin{array}{c}a \\ a^\#\end{array}\right].
\]
Furthermore, $[V,z] = [z^*,V]^*$ and  hence,
\begin{eqnarray}
\label{VzzV}
%\lefteqn{
[V,z] [z^*,V] =%}\nonumber \\&& 
4\left[\begin{array}{c}a \\ a^\#\end{array}\right]^\dagger PJ 
\Sigma \tilde E^T \tilde E^\# \Sigma
JP
\left[\begin{array}{c}a \\ a^\#\end{array}\right].
\end{eqnarray}
Also, we can write
\begin{equation}
\label{zz}
zz^* = \left[\begin{array}{c}a \\ a^\#\end{array}\right]^\dagger
\Sigma \tilde E^T \tilde E^\# \Sigma
\left[\begin{array}{c}a \\ a^\#\end{array}\right].
\end{equation}

Hence using Lemma \ref{L2}, we obtain
\begin{eqnarray}
\label{lyap_ineq3}
&&-\imath[V,\frac{1}{2}\left[\begin{array}{cc}a^\dagger &
      a^T\end{array}\right]M
\left[\begin{array}{c}a \\ a^\#\end{array}\right]]\nonumber \\
&&+ \frac{1}{2}L_a^\dagger[V,L_a]+\frac{1}{2}[L_a^\dagger,V]L_a
+ [V,z][z^*,V]
+\frac{zz^*}{\gamma^2}
 \nonumber \\
&=& \left[\begin{array}{c}a \\ a^\#\end{array}\right]^\dagger\left(\begin{array}{c}
F^\dagger P + P F\\ 
+4 PJ\Sigma \tilde E^T \tilde E^\# \Sigma JP \\
+ \frac{1}{ \gamma^2}\Sigma \tilde E^T \tilde E^\# \Sigma\\
\end{array}\right)\left[\begin{array}{c}a \\
a^\#\end{array}\right]\nonumber \\
&&+\tr\left(PJN_a^\dagger\left[\begin{array}{cc}I & 0 \\ 0 & 0 \end{array}\right]N_aJ\right)
\end{eqnarray}
where $F = -\imath JM-\frac{1}{2}JN_a^\dagger J N_a$. 

We now observe that using the  strict bounded real lemma, (\ref{Hurwitz1}) and (\ref{Hinfbound1}) imply  that the matrix inequality 
\begin{equation}
\label{QMI2}
F^\dagger P + P F 
+4 PJ\Sigma \tilde E^T \tilde E^\# \Sigma JP 
+ \frac{1}{ \gamma^2}\Sigma \tilde E^T \tilde E^\# \Sigma
 < 0.
\end{equation}
will have a solution $P > 0$ of the form (\ref{Pform}); e.g., see \cite{ZDG96,MaP4}.  This matrix $P$ defines a corresponding operator $V \in \mathcal{P}_1$ as in (\ref{quadV}). From this, it follows using (\ref{lyap_ineq3}) that there exists a constant $\delta_0 > 0$ such that 
\begin{eqnarray}
\label{dissip1a}
&&-\imath[V,\frac{1}{2}\left[\begin{array}{cc}a^\dagger &
      a^T\end{array}\right]M
\left[\begin{array}{c}a \\ a^\#\end{array}\right]]\nonumber \\
&&+ \frac{1}{2}L_a^\dagger[V,L_a]+\frac{1}{2}[L_a^\dagger,V]L_a
+ [V,z][z^*,V]
\nonumber \\&&
+ \frac{1}{ \gamma^2}zz^* 
+ \delta_0 \left[\begin{array}{c}a \\ a^\#\end{array}\right]^\dagger
\left[\begin{array}{c}a \\a^\#\end{array}\right]
\leq \tilde \lambda.
%\nonumber \\
\end{eqnarray}
with 
\[
\tilde \lambda = \tr\left(PJN_a^\dagger\left[\begin{array}{cc}I & 0 \\ 0 & 0 \end{array}\right]N_aJ\right) \geq 0.
\]
Also,  Lemma \ref{LB} and the fact that $V$ commutes with $L_b$ implies 
\begin{eqnarray}
\label{ineq1a}
\mathcal{G}(V) &\stackrel{\Delta}{=}& -\imath[V,H] + \frac{1}{2}L^\dagger[V,L]+\frac{1}{2}[L^\dagger,V]L\nonumber \\
 &=& -\imath[V,f(b,b^\#,z,z^*)]\nonumber \\
&&-\imath[V,\frac{1}{2}\left[\begin{array}{cc}a^\dagger &
      a^T\end{array}\right]M
\left[\begin{array}{c}a \\ a^\#\end{array}\right]]\nonumber \\
&&+ \frac{1}{2}L_a^\dagger[V,L_a]+\frac{1}{2}[L_a^\dagger,V]L_a\nonumber \\
&=&-\imath[V,\frac{1}{2}\left[\begin{array}{cc}a^\dagger &
      a^T\end{array}\right]M
\left[\begin{array}{c}a \\ a^\#\end{array}\right]]\nonumber \\
&&+ \frac{1}{2}L_a^\dagger[V,L_a]+\frac{1}{2}[L_a^\dagger,V]L_a\nonumber \\
&&-\imath[V,z]w_{1}^*+\imath w_{1}[z^*,V]
%\nonumber \\&&
-\frac{1}{2}\imath\mu w_{2}^*+\frac{1}{2}\imath  w_{2}\mu^*. \nonumber \\
\end{eqnarray}
Here, $\mathcal{G}(\cdot)$ is the generator associated with the quantum system defined by $(S,L,H)$; e.g., see \cite{JG10}.  
Furthermore, 
\begin{eqnarray*}
0 &\leq& \left([V,z]- \imath w_{1}\right)
\left([V,z]- \imath w_{1}\right)^*\nonumber \\
&=&  [V,z][z^*,V]+\imath[V,z]w_{1}^*%\nonumber \\&&
-\imath  w_{1}[z^*,V]+ w_{1} w_{1}^*
\end{eqnarray*}
and hence
\begin{eqnarray}
\label{ineq3a}
% \lefteqn{
-\imath[V,z]w_{1}^*+\imath w_{1}[z^*,V]
%}\nonumber \\
&\leq& [V,z][z^*,V]+ w_{1} w_{1}^*.\nonumber \\
\end{eqnarray}

Also, 
\begin{eqnarray*}
0 &\leq&  \left(\frac{1}{2}\mu- \imath w_{2}\right)
\left(\frac{1}{2}\mu- \imath w_{2}\right)^*\nonumber \\
&=&  \frac{1}{4}  \mu \mu^*-\frac{\imath}{2}  w_{2}\mu^*+\frac{\imath}{2} \mu w_{2}^*
%\nonumber \\&&
+  w_{2}  w_{2}^*
\end{eqnarray*}
and hence
\begin{eqnarray}
\label{ineq3b}
%\lefteqn{
\frac{\imath}{2} w_{2}\mu^*-\frac{\imath}{2} \mu w_{2}^*
%}\nonumber \\
 &\leq& \frac{1}{4} \mu\mu^*
%\nonumber \\&&
+ w_{2} w_{2}^*.%\nonumber \\
\end{eqnarray}
Substituting (\ref{ineq3a}), (\ref{ineq3b}),  into (\ref{ineq1a}), it follows that
\begin{eqnarray}
\label{ineq2a}
\lefteqn{\mathcal{G}(V)+ \delta_0 \left[\begin{array}{c}a \\ a^\#\end{array}\right]^\dagger
\left[\begin{array}{c}a \\a^\#\end{array}\right]}\nonumber \\
 &\leq &  -\imath[V,\frac{1}{2}\left[\begin{array}{cc}a^\dagger &
      a^T\end{array}\right]M
\left[\begin{array}{c}a \\ a^\#\end{array}\right]]\nonumber \\
&& + \frac{1}{2}L_a^\dagger[V,L_a]+\frac{1}{2}[L_a^\dagger,V]L_a\nonumber \\
&&+  [V,z][z^*,V]+\frac{1}{\gamma^2} z z^*+ \frac{1}{4}\mu\mu^*
\nonumber \\&&
+ w_{1} w_{1}^* - \frac{1}{\gamma^2} z z^*
+w_{2} w_{2}^*.
\end{eqnarray}
Then it follows from (\ref{dissip1a}) that 
\begin{eqnarray}
\label{ineq4}
\lefteqn{\mathcal{G}(V)+ \delta_0 \left[\begin{array}{c}a \\ a^\#\end{array}\right]^\dagger
\left[\begin{array}{c}a \\a^\#\end{array}\right]}\nonumber \\
& \leq &\tilde \lambda +  \frac{1}{4}\mu\mu^*+ w_{1} w_{1}^* - \frac{1}{\gamma^2} z z^*+w_{2} w_{2}^*. 
\end{eqnarray}

Now using a similar argument to that used in the proof of Lemma 3.4 in \cite{JG10}, it follows from (\ref{ineq4}) that given any $T > 0$,
\begin{eqnarray*}
\lefteqn{\left<V(T)\right> - V(0)}\nonumber \\
 &=& \int_0^T \left< \mathcal{G}(V(t))\right>dt \nonumber \\
&\leq & \left(\lambda +  \frac{1}{4}\mu\mu^*\right)T \nonumber \\
&&- \delta_0\int_0^T \left<\left[\begin{array}{c}a(t) \\ a(t)^\#\end{array}\right]^\dagger
\left[\begin{array}{c}a(t) \\a(t)^\#\end{array}\right]\right>dt \nonumber \\
&&+\int_0^T\left(\left<w_1(t) w_1(t)^*\right>-  \frac{1}{\gamma^2}\left<z(t) z(t)^*\right>\right)dt \nonumber \\
&&+ \int_0^T\left<w_2(t) w_2(t)^*\right>dt.
\end{eqnarray*}
However, $\left<V(T)\right> \geq 0$ since $P > 0$ and therefore, we can write
\begin{eqnarray*}
\lefteqn{\frac{\delta_0}{T}\int_0^T \left<\left[\begin{array}{c}a(t) \\ a(t)^\#\end{array}\right]^\dagger
\left[\begin{array}{c}a(t) \\a(t)^\#\end{array}\right]\right>dt}\nonumber \\
 &\leq& \frac{V(0)}{T} + \lambda +  \frac{1}{4}\mu\mu^* \nonumber \\
&&+\frac{1}{T}\int_0^T\left(\left<w_1(t) w_1(t)^*\right>-  \frac{1}{\gamma^2}\left<z(t) z(t)^*\right>\right)dt \nonumber \\
&&+ \frac{1}{T}\int_0^T\left<w_2(t) w_2(t)^*\right>dt.
\end{eqnarray*}
Taking the lim sup as $T \rightarrow \infty$ on both sides of this inequality and using (\ref{sector4a}), (\ref{sector4b}), it follows that
\begin{eqnarray*}
\lefteqn{\limsup_{T \rightarrow \infty}\frac{1}{T}\int_0^T \left<\left[\begin{array}{c}a(t) \\ a(t)^\#\end{array}\right]^\dagger
\left[\begin{array}{c}a(t) \\a(t)^\#\end{array}\right]\right>dt} \nonumber \\
&\leq & \frac{\lambda}{\delta_0} +  \frac{1}{4\delta_0}\mu\mu^* + \frac{\delta_1}{\delta_0} + \frac{\delta_2}{\delta_0}. 
\end{eqnarray*} 
 Hence, the condition (\ref{ms_stable0}) is satisfied with 
\[
c = \frac{\lambda}{\delta_0} +  \frac{1}{4\delta_0}\mu\mu^* + \frac{\delta_1}{\delta_0} + \frac{\delta_2}{\delta_0}
 \geq 0.
\] 
\hfill $\Box$

\section{Illustrative Example}
\label{sec:example}
\begin{figure}[hbp]
\psfrag{k1}{\color{red} $\kappa_1$}
\psfrag{k2}{\color{blue} $\kappa_2$}
\psfrag{chi2}{\small $\chi^{(2)}$}
\begin{center}
\includegraphics[width=8cm]{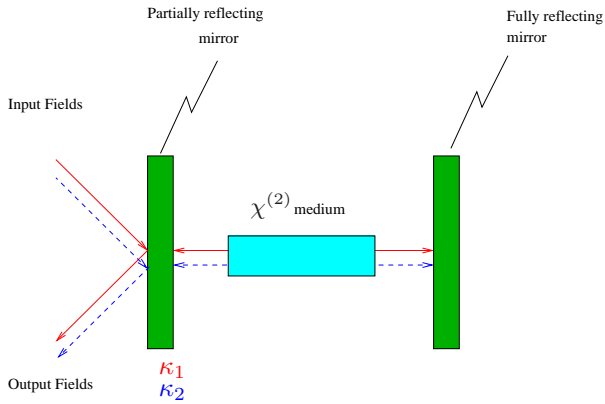}
\end{center}
\caption{Schematic diagram of an OPA system. Here, the red solid lines represent the fields at the fundamental frequency and the blue dashed lines represent the fields at the second harmonic frequency.}
\label{F1}
\end{figure}

In this section, we present an example to illustrate the theory developed in this paper. In this example, we consider the linearized model of an optical parametric amplifier (OPA); e.g., see \cite{BR04,SHHPJ2a}. An OPA consists of a $\chi^{(2)}$ optical medium contained in an optical cavity driven by coherent fields at a fundamental and second harmonic frequencies; e.g., see \cite{BR04,WM08}. The $\chi^{(2)}$  medium allows for coupling between the fundamental electromagnetic field and the second harmonic electromagnetic field.
The construction of an OPA is illustrated in Figure \ref{F1}. 
 
This quantum system is described by the triple $(S,L,H)$ where $S=I$, 
\[
L = \left[\begin{array}{c}
L_a\\L_b
\end{array}\right],~~L_a= \sqrt{\kappa_a} a,~~L_b= \sqrt{\kappa_b} b,
\]
 and $H = \imath \chi \left(2 \bar a b^*a+ \bar b^*a^2 - \bar b a^{*2} - 2 \bar a^* a^* b\right)$. Here, $a$ is the annihilation operator associated with the fundamental mode of the system and $b$  is the annihilation operator associated with the second harmonic mode of the system. Also, $\chi>0$ is a constant associated with the  $\chi^{(2)}$ optical medium, and  $\kappa_a > 0$ and $\kappa_b > 0$ are constants associated with the cavity mirror reflectivities at the fundamental and second harmonic frequencies respectively. Furthermore, $\bar a$ and $\bar b$ are complex constants representing the steady state values of the fundamental and second harmonic fields within the cavity.

This Hamiltonian can be regarded as being of the form (\ref{H1}) with 
\[
M = \left[\begin{array}{cc}0 & -\imath \chi \bar b \\
\imath \chi \bar b^* & 0
\end{array}\right],
\]
$f(b,b^\#,z,z^*) = 2 \imath \chi \left(\bar a b^*z-\bar a^* z^*b\right)$, and $z = a$. Also, we calculate
\[
N_a = \left[\begin{array}{cc} \sqrt{\kappa_a} & 0 \\ 0 & \sqrt{\kappa_a}\end{array}\right],~~
N_b = \left[\begin{array}{cc} \sqrt{\kappa_b} & 0 \\ 0 & \sqrt{\kappa_b}\end{array}\right].
\]
In order to apply Theorem \ref{T1} to this system, we first calculate the dynamics of the system uncertainty 
(\ref{uncert_dynamics}). Indeed, we calculate 
\begin{eqnarray*}
[b,f(b,b^\#,z,z^*)] &=& 2 \imath \chi\bar a z;\nonumber \\
JN_b^\dagger JN_b  &=& \left[\begin{array}{cc} \kappa_b & 0 \\ 0 & \kappa_b\end{array}\right];\nonumber\\
JN_b^\dagger J &=& \left[\begin{array}{cc} \sqrt{\kappa_b} & 0 \\ 0 & \sqrt{\kappa_b}\end{array}\right];\nonumber\\
w_1 &=& \frac{\partial f(b,b^\#,z,z^*)^*}{\partial z} = - 2 \imath \chi\bar a^* b;\nonumber\\
w_2 &=& 0.
\end{eqnarray*}
 Hence, (\ref{uncert_dynamics}) implies 
\begin{eqnarray}
\label{b-system}
d b &=& -\frac{\kappa_b}{2}b dt + 2\chi\bar a zdt -\sqrt{\kappa_b}d\mathcal{B};\nonumber \\
w_1 &=& - 2 \imath \chi\bar a^* b.
\end{eqnarray}
In this case, the dynamics for $b^*$ are decoupled from these dynamics and need not be considered. Setting the noise input to zero in the system (\ref{b-system}), we calculate the transfer function from $z$ to $w_1$ to be 
\[
G(s) = -\frac{4 \imath \chi^2 \bar a\bar a^*}{s+ \frac{\kappa_b}{2}}.
\]
This transfer function is stable and has $H^\infty$ norm 
$\|G(s)\|_\infty = |G(0)| = \frac{8 \chi^2 \bar a\bar a^*}{\kappa_b}$. Also, if we set the input $z$ to zero in (\ref{b-system}), we can calculate the steady state covariance of $w_1(t)$ as
\[
\limsup_{T\rightarrow \infty} \frac{1}{T}\int_0^T\left<w_1(t) w_1(t)^*\right>dt = 4\chi^2 \bar a\bar a^*.
\]
Then, since the system (\ref{b-system}) is linear, it follows that the condition (\ref{sector4a}) will be satisfied with 
\begin{equation}
\label{gamma}
\gamma = \frac{\kappa_b}{8 \chi^2 \bar a\bar a^*}
\end{equation}
 and $\delta_1 = 4\chi^2 \bar a\bar a^*$. Also, since $w_2 = 0$, it follows that condition (\ref{sector4b}) is satisfied with $\delta_2 =0$. 

We now calculate the matrices  $F$ and $\tilde E$,  and the transfer function $H(s) = \tilde E^\# \Sigma\left(sI -F\right)^{-1}J\Sigma \tilde E^T$ in order to check  conditions (\ref{Hurwitz1}) and (\ref{Hinfbound1}). Indeed, we calculate
\[
F= \left[\begin{array}{cc}
-\frac{\kappa_a}{2} & - \chi \bar b\\- \chi \bar b^* &  -\frac{\kappa_a}{2}
\end{array}\right],~~\tilde E = \left[\begin{array}{cc}1 & 0 \end{array}\right],
\]
and 
\[
H(s) 
= \frac{-\left(s+\frac{\kappa_a}{2}\right)}{s^2 + \kappa_a s + \frac{\kappa_a^2}{4}-\chi^2 \bar b \bar b^*}.
\]
It is straightforward to verify that the matrix $F$ is Hurwitz if and only if 
\begin{equation}
\label{Hurwitz_cond}
\kappa_a > 2 \chi |\bar b|.
\end{equation} 
Also,  $H(s)$ has a real zero at $s = - \frac{\kappa_a}{2}$ and real poles at $s = - \frac{\kappa_a}{2} \pm \chi |b|$. From this, it follows that $H^\infty$ norm of $H(s)$ is given by 
\[
\|H(s)\|_\infty = |H(0)| = \frac{2 \kappa_a}{\kappa_a^2 - 4 \chi^2 \bar b \bar b^*}.
\]
Hence, it follows using (\ref{gamma}) that condition (\ref{Hinfbound1}) is satisfied if and only if 
\begin{eqnarray*}
&&\frac{2 \kappa_a}{\kappa_a^2 - 4 \chi^2 \bar b \bar b^*} < \frac{\kappa_b}{16 \chi^2 \bar a\bar a^*}\\
&\Leftrightarrow & 32 \frac{\kappa_a}{\kappa_b}\chi^2 \bar a\bar a^* < \kappa_a^2 - 4 \chi^2 \bar b \bar b^*\\
&\Leftrightarrow & 4\chi^2 \left(8 \frac{\kappa_a}{\kappa_b}\bar a\bar a^*+\bar b \bar b^*\right) < \kappa_a^2.
\end{eqnarray*}
Note, that if this condition is satisfied, then it immediately follows that the condition (\ref{Hurwitz_cond}) will be satisfied and hence, the matrix $F$ will be Hurwitz. Hence using Theorem \ref{T1}, we can conclude that if this condition is satisfied then the OPA system will be robustly mean square stable. 

\section{Conclusions}
\label{sec:conclusions}
In this paper, we have extended the robust stability result of \cite {PUJ1a}  to the case of nonlinear dynamic uncertainties described in terms of a stochastic integral quadratic constraint. This also led to a robust stability condition in the form of a small gain condition. This condition was then applied the robust stability
analysis of a quantum system model for an OPA and  a stability condition for this system was obtained.
%\bibliography{/home/irp/Bibliog/irpnew}
%\bibliography{/home/s8504138/Bibliog/irpnew}  
%\bibliographystyle{IEEEtran}

\end{document}